\documentstyle[12pt]{article}
\newcommand{\beq}{\begin{equation}}
\newcommand{\eeq}{\end{equation}}
\newcommand{\beqa}{\[}
\newcommand{\eeqa}{\]}
\newcommand{\bea}{\begin{eqnarray}}
\newcommand{\eea}{\end{eqnarray}}  
\begin{document}
\begin{flushright}
IFIC/02-03
\end{flushright}
\vskip 2.0cm
\begin{center}
\centerline{\Large\bf Vacuum Polarization Effects from Fermion}
\vskip 0.5cm
\centerline{\Large\bf  Zero-Point Energy} 
\vspace{1.5cm}
  
J. L. Tomazelli\footnote{On leave from Departamento de F\'{\i}sica e 
Qu\'{\i}mica, UNESP, Campus de Guaratinguet\'{a}, Brazil}
\\
\vspace{0.5cm}

{\em Departament de F\'{\i}sica Te\`{o}rica, IFIC-CSIC, Universitat de
Val\`{e}ncia \\
46100 Burjassot, Val\`{e}ncia, Spain}
\vspace{1.5cm}
\end{center} 

\begin{abstract}

The influence of an external electromagnetic field on the vacuum structure
of a quantized Dirac field is investigated by considering the quantum 
corrections to classical Maxwell's Lagrangian density induced by fluctuations
of the non-perturbative vacuum. Effective Lagrangian densities for Maxwell's
theory in (3+1) and (2+1) dimensions are derived from the vacuum zero-point 
energy of the fermion field in the context of a consistent 
Pauli-Villars-Rayski subtraction scheme, recovering Euler-Kockel-Heisenberg
and Maxwell-Chern-Simons effective theories. Effective Scalar Quantum Electrodynamics 
as well as low temperature effects in both spinor and scalar theories are also discussed.    
 
\end{abstract}  
\newpage

\section{Introduction}

In the early 30's, one of the more striking results from Dirac's positron theory,
in addition to the possibility of converting electromagnetic energy into matter
through electron-positron pair production, in sufficiently strong electromagnetic 
fields, was the scattering of quantized electromagnetic radiation by other photons
or an external electromagnetic field, the latter process known as Delbr\"uck
scattering. These nonlinear effects, which were not predicted by classical 
electromagnetic theory, were studied by Euler and Kockel and Heisenberg and 
Euler${}^{[1]}$, who derived an effective Lagrangian density for such an extended
version of Maxwell's theory, induced by an external static and homogeneous
electromagnetic field. Their work was refined by Weisskopf${}^{[2]}$, the first to
introduce the idea of charge renormalization, in the scope of the 
so-called physics of subtractions. 

Although Weisskopf approach was not manifestly Lorentz invariant, he came up with
an effective Lagrangian wich is equivalent to the Lorentz and gauge invariant
Lagrangian density derived by Schwinger${}^{[3]}$ some years latter. However, along 
his calculations, Weisskopf dealt with divergent sums and momentum integrals, 
without any regularization prescription to give them a precise meaning from the
mathematical standpoint. Only in the late 40's, short before the previously 
mentioned Schwinger's approach, the physics of subtractions would be practiced 
with proper rigour in the works by Rayski and Pauli and Villars${}^{[4]}$, who 
first introduced a Lorentz and gauge invariant regularization prescription to 
tackle the singular Green's functions of quantum electrodynamics (QED).  
       
Along the last two decades Effective theories have deserved special attention in 
different contexts, ranging from quantum electrodynamics in external fields${}^{[5,6]}$, 
the linear sigma model${}^{[7]}$ and the Nambu-Jona-Lasinio${}^{[8]}$ in hadron 
physics to models for low-energy quantum chromodynamics (QCD)${}^{[9]}$ and gauge 
field theories in (2+1) space-time dimensions, in particular, quantum electrodynamics 
(QED$_3$)${}^{[10]}$ and the non-renormalizable gauged Thirring model${}^{[11]}$, 
either at zero or at finite\- temperature${}^{[12]}$. 

The aim of this paper is to rederive Euler-Kockel-Heisenberg effective Lagrangian
for QED and QED$_3$ following the approch of Berestetski, Lifshitz and 
Pitayeviski${}^{[13]}$, which is inspired by Weisskopf's essential ideas, 
employing Pauli-Villars-Rayski regularization prescription in order to keep under 
control divergent quantities which calls for a consistent subtraction procedure.
The next section presents the formalism and the discussion of the necessity of 
introducing auxiliary regulator masses into the subtracting sheme of divergences 
in the fermion zero-point energy for QED in the presence of a constant and uniform
magnetic field. In section 3, the (2+1) dimensional case is studied, paying 
attention to the conditions that must be imposed on the regulator masses in order 
to eliminate the divergences of the theory and discuss the problem of parity 
violation in the effective Lagrangian density. The same analysis is also 
applied to scalar quantum electrodynamics (s-QED), in section 4. The last section 
is devoted to some concluding remarks.
               
\section{QED Effective Lagrangian in (3+1) Dimensions}

In order to determine the effective Lagrangian density for the electromagnetic field
from the zero-point energy of matter, which in this case is represented by the
electron-positron field, Dirac Hamiltonian must be written in Fock space as
\beq
{\cal H}=\sum_{{\bf p},\,\sigma}\epsilon_{{\bf p}\sigma}
(a^\dagger_{{\bf p}\sigma}a_{{\bf p}\sigma}+
b^\dagger_{{\bf p}\sigma}b_{{\bf p}\sigma})+\varepsilon_0\,\,,  
\eeq
where
\bea
\varepsilon_0 &=&\langle{\cal H}\rangle 
=-\sum_{{\bf p},\,\sigma}\epsilon_{{\bf p}\sigma}^{(-)} \nonumber \\
&=&\sum_{{\bf p},\,\sigma}\int\,d^3x\,\psi_{{\bf p}\sigma}^{(-)^*}i
\frac{\partial}{\partial t}\psi_{{\bf p}\sigma}^{(-)}
\eea
is the vacuum zero-point energy. Here $\psi_{{\bf p}\sigma}^{(-)}$ are the negative
frequence solutions of Dirac equation in the presence of the electromagnetic field, 
normalized in a unit volume, so that $\varepsilon_0$ is in fact a {\em scalar
density}. The expectation value of the potential energy of an electron in 
these negative energy states is
\bea
U_0&=&\sum_{{\bf p},\,\sigma}\int\,d^3x\,\psi_{{\bf p}\sigma}^{(-)^*}e\phi
\psi_{{\bf p}\sigma}^{(-)} \nonumber \\
&=&{\bf E}.\sum_{{\bf p},\,\sigma}\int\,d^3x\,\psi_{{\bf p}\sigma}^{(-)^*} 
\frac{\partial {\cal H}}{\partial {\bf E}}\psi_{{\bf p}\sigma}^{(-)} 
={\bf E}.\frac{\partial\varepsilon_0}{\partial {\bf E}}\,,
\eea
where $\phi=-{\bf E}.{\bf r}$ is the scalar potential of a uniform electric field. 

The above quantity must be subtracted from $\varepsilon_0$, since the physical 
vacuum is expected to have zero charge. Therefore, the total shift in the vacuum 
energy density induced by an external uniform electromagnetic field is
\beq
\delta W=\left(\varepsilon_0-{\bf E}.\frac{\partial\varepsilon_0}{\partial {\bf E}}
\right)-\left(\varepsilon_0-{\bf E}.\frac{\partial\varepsilon_0}{\partial {\bf E}}
\right)_{{\bf E}={\bf H}=0}\,\,.
\eeq
This represents an additional contribution to the total electromagnetic e\-nergy
density $W$ due to vacuum polarization effects. This energy is related to the 
effective Lagrangian density ${\cal L}_{\rm eff}$ of the electromagnetic field by 
the transformation
\beq
W={\bf E}.\frac{\partial {\cal L_{\rm eff}}}{\partial {\bf E}}-
{\cal L_{\rm eff}}\,,
\eeq
so that its variation can be compared with expression (4). As a result,
\beq
\delta{\cal L}={\cal L}_{\rm eff}-{\cal L}_0=-[\varepsilon_0-
(\varepsilon_0)_{{\bf E}={\bf H}=0}]\,,
\eeq
where ${\cal L}_0$ is the Lagrangian density for the applied external 
electromagnetic field.

The negative energy levels of an electron with charge $-|e|$ in a constant and 
uniform magnetic field $H_z=-H $ are
\beq
-\epsilon_{{\bf p}\sigma}^{(-)}=-\sqrt{m^2+(2n+1-\sigma)|e|H+p_z^2}\,,
\eeq
where $n=0,1,2,\dots$ and $\sigma=\pm 1$. The sum over the $z$-components of 
the electron momenta in (2) will be evaluated in the continnum by considering the
corresponding density of momentum states 
\beqa
\frac{|e|H}{2\pi}\frac{dp_z}{2\pi}\,.
\eeqa
Thus, the negative vacuum zero-point energy can be written as
\bea
-\varepsilon_0&=&\frac{|e|H}{(2\pi)^2}\int_{-\infty}^{\infty}\,dp_z
\sum_{n=0}^{\infty}\sum_{\sigma=\pm 1}\epsilon_{n,\sigma}(p_z) \nonumber \\
&=&\frac{|e|H}{(2\pi)^2}\int_{-\infty}^{\infty}\,dp_z
\left\{\sqrt{m^2+p_z^2}+2\sum_{n=1}^{\infty}
\sqrt{m^2+2|e|Hn+p_z^2}\right\}\,.
\eea
The above momentum integrals are quadratically divergent at the ultraviolet. In
order to extract a finite result for $\delta{\cal L}$, a consistent subtration
procedure is necessary, starting from a mathematical meaningful expression 
for the zero-point energy, i.e., $\varepsilon_0$ must be {\em regularized}.
This can be achived in the sense of Pauli-Villars-Rayski regularization scheme, 
by introducing auxiliary masses of ficticious {\em regulator fields}, satisfyng 
suitable conditions that eliminate the divergences of the original theory, which 
is ultimatelly recovered by letting the regulator masses go to infinite at the 
end of calculations.

The first step is to analyze the conditions fulfilled by the auxiliary masses 
according to the degree of singularity of the generalized function
\beq
{\cal D}(m^2)=\int_{-\infty}^{\infty}\,dp\,\sqrt{m^2+p^2}\,.
\eeq
This may be interpreted in the sense of its analytical continuation
\beq
{\cal D}_\delta(m^2)=\lim_{\delta\rightarrow 0}\,\int_{-\infty}^{\infty}\,dp
\,(m^2+p^2)^{\frac{1}{2}-\delta}
\eeq
which can be differentiated with respect to the parameter $m^2$ {\em before} 
taking the limit $\delta\rightarrow 0$. The asymptotic expansion of integral 
(9) for $p^2\gg m^2$,
\beqa
{\cal D}(m^2)\sim\int\,dp\,p\,(1+\frac{m^2}{2p^2}) + ({\rm regular\,\,
terms})\,, 
\eeqa 
shows that conditions
\beq
\sum_{i=0}^{N}c_i=0 \,,\,\,\sum_{i=0}^{N}c_im_i^2=0
\eeq
must be imposed on the regulated expression
\beq
{\cal D}^R(m^2)=\sum_{i=0}^{N}c_i{\cal D}(m_i^2)
\eeq 
in order to eliminate the quadratic and logarithmic divergences, 
respectively. In the above expressions, $N$ is the total number of 
regulators, $c_0=1$ and $m_0=m$ is the electron mass.  

Instead of considering the regulated expression $\varepsilon_0^R$ 
for the zero-point e\-nergy (8), its easier to cope with the function
\beq
\Phi^R(H)\equiv-\frac{\partial^2\varepsilon_0^R}{\partial (m^2)^2} 
=\sum_ic_i\Phi_i(H) \,,
\eeq
where
\bea
\Phi_i(H)&=&-\frac{|e|H}{2(2\pi)^2}\int_0^{\infty}\,dp_z\left\{
(m_i^2+p_z^2)^{-3/2-\delta}\right. \nonumber \\
&+&\left.2\sum_{n=1}^{\infty}
(m_i^2+2|e|Hn+p_z^2)^{-3/2-\delta}\right\} 
\eea
and perform the finite momentum integrals for $\delta\rightarrow 0$. 
As a result,
\beq
\Phi_i(H)=-\frac{|e|H}{2(2\pi)^2}\left\{\frac{1}{m_i^2}+
2\sum_{n=1}^{\infty}\frac{1}{m_i^2+2|e|Hn}\right\}\,.
\eeq
The remaining sum still carries a divergence and can be evaluated after making 
use of the gamma function integral representation
\beq
\frac{1}{A^{1+\delta}}=\frac{1}{\Gamma(1+\delta)}\int_{0+}^{\infty}
\,d\eta\,\eta^{\delta}e^{-A\eta}\,,
\eeq
valid for $\delta>-1$. It follows that
\bea
\Phi_i(H)&=&-\frac{|e|H}{8\pi^2}\frac{1}{\Gamma(1+\delta)}\int_{0+}^{\infty}
\,d\eta\,\eta^{\delta}e^{-m_i^2\eta}\left[2\sum_{n=0}^{\infty}
e^{-2|e|Hn\eta}-1\right] \nonumber \\ 
&=&-\frac{|e|H}{8\pi^2}\frac{1}{\Gamma(1+\delta)}\int_{0+}^{\infty}\,d\eta\,
\eta^{\delta}e^{-m_i^2\eta}{\rm coth}(|e|H\eta) \,. 
\eea
Two successive integrations with respect to $m_i^2$ give, for 
$\delta\rightarrow 0$,
\beq
-\varepsilon_{0\,i}=-\frac{|e|H}{8\pi^2}\int_{0+}^{\infty}\,\frac{d\eta}
{\eta^2}\,e^{-m_i^2\eta}{\rm coth}(|e|H\eta) + C^{(0)}(H) + C^{(2)}(H)m_i^2
\,,
\eeq
where $C^{(0)}$ and $C^{(2)}$ are eventually infinite constants which do not 
depend on $m_i^2$ and, therefore, can be absorbed into the coefficients 
$c_i$'s and eliminated in virtue of conditions (11). 

Hence, from equation (6), the regularized correction to the 
electromagnetic Lagrangian density reads
\beq
\delta{\cal L}^R=-\frac{1}{8\pi^2}\sum_ic_i\int_{0+}^{\infty}\,
d\eta\,\frac{e^{-m_i^2\eta}}{\eta^3}[|e|H\eta{\rm coth}(|e|H\eta)-1]\,.
\eeq
For $H\ll m_i^2$, individual contributions to the above sum have the form
$m_i^4F\left(\frac{H^2}{m_i^4}\right)$, where $F$ is an adimensional 
function, whose power series expansion does not contain a divergent $H$ 
independent term. The remaining divergent terms, proportional to $H^2$, 
are eliminated from the sum by the first condition in (11). For this 
purpose, it's sufficient to introduce two auxiliary masses $M_1$ and $M_2$
and $c_1=c_2=-1/2$, so that conditions (11) are always satisfied, even 
for large regulator masses. This can be achieved by choosing $M_1$ real 
and $M_2$ imaginary. The regulator contributions
\beq
c_1\delta{\cal L}_1+c_2\delta{\cal L}=\frac{e^2H^2}{3(8\pi^2)}
\int_{0+}^{\infty}\,d\eta\,\frac{e^{-\eta}}{\eta}
\eeq
corresponds to the well-known charge renormalization counterterm.

Finally, letting $|M_1|\,,\,|M_2|\rightarrow\infty$ gives the expected 
finite result  
\beq
\delta{\cal L}=\frac{m^4}{8\pi^2}\int_{0+}^{\infty}\,d\eta\,
\frac{e^{-\eta}}{\eta^3}\left\{-\eta b\,{\rm coth}(\eta b)+1
-\frac{1}{3}b^2\eta^2\right\}\,,
\eeq
where $b=|e|H/m^2$. It can be shown${}^{[13]}$ by dimensional analysis 
and from the invariants ${\bf a}^2-{\bf b}^2$ and ${\bf a}.{\bf b}$ of 
the electromagnetic theory that, in the presence of constant and 
uniform magnetic and electric fields, parallel to the $z$-direction,
\beq
\delta{\cal L}=\frac{m^4}{8\pi^2}\int_{0+}^{\infty}\,d\eta\,
\frac{e^{-\eta}}{\eta^3}\left\{-b\eta{\rm coth}(\eta b)
a\eta{\rm cotg}(\eta a)+1-\frac{1}{3}(b^2-a^2)\eta^2\right\} \,, 
\eeq
where ${\bf b}=|e|{\bf H}/m^2$ and ${\bf a}=|e|{\bf E}/m^2$, a result
derived by Schwinger${}^{[3]}$ using the Fock proper-time 
method${}^{[14]}$.

Note that both integrals in expression (8) have the same 
degree of singularity and were regularized as a whole divergent 
object. However, if they were treated independently, the additional 
constraint
\beq
\sum_{i=0}^{N}c_im_i^2\log(m_i^2)=0
\eeq 
would have to be imposed on each regularized integral in order to 
keep them finite, removing terms linear in $H$, which are not 
symmetric under reflections $H\rightarrow -H$. In fact, two successive
integrations of
\bea
\Phi^{(0)\,R}&\equiv&\sum_ic_i\frac{1}{(m_i^2)^{1+\delta}} \nonumber \\
&=&\frac{1}{\Gamma(1+\delta)}\sum_ic_i\int_{0+}^{\infty}\,d\eta\, 
\eta^{\delta}e^{-m_i^2\eta} \nonumber
\eea
with respect to $m^2$ give
\bea
\varepsilon_0^{(0)\,R}&=&\frac{1}{\Gamma(1+\delta)}\sum_ic_i
\int_{0+}^{\infty}\,d\eta\,\eta^{\delta-2}e^{-m_i^2\eta} \nonumber \\
&=&\frac{\Gamma(\delta-1)}{\Gamma(\delta+1)}\sum_ic_i(m_i^2)^{1-\delta}
=\sum_ic_im_i^2\log(m_i^2)+{\cal O}(\delta) \nonumber \,, 
\eea
where conditions (11) have been used. In this way, the final result for
$\delta{\cal L}$ would be the same only if finite counterterms were 
added in order to preserve parity invariance. A quite analogous 
situation is found in scalar QED, and will be examined later.   

\section{QED Effective Lagrangian in (2+1) Dimensions}

In (2+1) dimensions the electron energy levels in the presence of a 
constant and uniform magnetic field, derived from the 3-potential
$A_{\mu}=(0,0,xH)$, are given by
\beq
|\epsilon_{n\sigma}|=\sqrt{m^2+(2n+1-\sigma)|e|H}\,.
\eeq
This result is obtained by solving Dirac equation in the minimal 
representation of Clifford algebra, choosing $\gamma_0=\sigma_z$, 
$\gamma_1=i\sigma_x$ and $\gamma_2=i\sigma_y$. In QED$_3$, the 
coupling constant $e$ and the components $A_{\mu}$ have the same 
dimensionality [$e$]=[$A$]=[$m$]$^{1/2}$, so that [$H$]=[$m$]$^{3/2}$. 
This is of fundamental importance for the analysis of the functional 
dependence of $\delta{\cal L}$ on $H$ and the fermion mass. 

The vacuum zero-point energy follows from (2) and (24),   
\beq
-\varepsilon_0=\frac{|e|H}{2\pi}\left[|m|+2\sum_{n=1}^{\infty}
\sqrt{m^2+2|e|Hn}\right]=\varepsilon_0^{(0)}+\varepsilon_0^{(1)}\,,
\eeq
where $|e|H/(2\pi)$ amounts for the density of states in the 
(2+1)-dimensional phase space. In contrast with the four-dimensional 
case, the odd-parity term $\varepsilon_0^{(0)}$ is finite and does not 
need to be regularized. The divergent contribution must be replaced by
its regularized counterpart
\beq
\varepsilon_0^{(1)\,R}=\sum_ic_i\varepsilon_{0\,i}^{(1)} \,,
\eeq 
where each term, when differentiated with respect to $m_i^2$, yields a function
\bea
\Phi^{(1)}_i(H)&=&\frac{|e|H}{(2\pi)}\sum_{n=1}^{\infty}
(m_i^2+2|e|Hn)^{-1/2-\delta} \nonumber \\
&=&\frac{|e|H}{2\pi\Gamma(1/2)}\int_{0+}^{\infty}\,d\eta\,
\eta^{-1/2+\delta}e^{-m_i^2\eta}\sum_{n=1}^{\infty}e^{-2|e|Hn\eta} 
\nonumber \\
&=&\frac{|e|H}{4\pi\sqrt{\pi}}\int_{0+}^{\infty}\,d\eta\,
\eta^{-1/2+\delta}\frac{e^{-(m_i^2+|e|H)\eta}}{{\rm sinh}(|e|H\eta)} \,,
\eea
which admits the integral representation (16). After integration of
(27) with respect to $m^2$, the regularized zero-point energy reads 
\beq
-\varepsilon_{0}^R=\frac{|e|H|m|}{2\pi}
-\frac{e|H|}{4\pi\sqrt{\pi}}\sum_ic_i\int_{0+}^{\infty}\,d\eta\,
\eta^{-3/2}\frac{e^{-(m_i^2+|e|H)\eta}}{{\rm sinh}(|e|H\eta)} \,, 
\eeq 
where the integration constant has been suppressed by the first 
condition in (11), assumed to be valid also in the present case, so that
any contribution to the regularized expression $\delta{\cal L}^R$ has 
the form $m_i^3F\left(|e|H/m_i^2\right)$, where $F$ is an adimensional
function whose $H$ series expansion does not allow even powers of the 
field masses to appear in the corrected electromagnetic Lagrangian 
density. On the other hand, this functional dependence of $F$ on the 
magnetic field $H$ indicates that the effective Lagrangian density for 
the (2+1)-dimensional electromagnetic field might not be invariant under 
parity transformations. This occurs in QED$_3$, as a consequence of a
non-invariant fermion mass term in Dirac Lagrangian density, which is
responsible for the dynamically generated odd-parity term in the
renormalized electromagnetic sector${}^{[15]}$.

Thus, from (6),
\beq
\delta{\cal L}^R=\frac{|e|H|m|}{2\pi}-\frac{1}{4\pi\sqrt{\pi}}
\sum_ic_i\int_{0+}^{\infty}\,\frac{d\eta}{\eta^{5/2}}\,e^{-m_i^2\eta}
\left[\frac{|e|H\eta\,e^{-|e|H\eta}}{{\rm sinh}(|e|H\eta)}-1\right] \,.
\eeq
For a weak applied magnetic field, the terms proportional to $H$ in the 
power series expansion of the function in square brackets give rise to 
contributions 
\beqa
\frac{|e|H|m_i|}{4\pi\sqrt{\pi}}\int_{0+}^{\infty}
\frac{d\eta}{\eta^{3/2}}\,e^{-\eta} \,,
\eeqa 
linear in the field masses, which can be eliminated by the subsidiary 
condition
\beq
\sum_{i=0}^{N}c_i|m_i|=0 \,.
\eeq
This can be accomplished by introducing again two regulator masses 
$M_1$ and $M_2$, exactly as in the four-dimensional case, with 
$c_1=c_2=-1/2$. The remaining odd-parity contributions are finite.

Hence, for $|M_1|\,,\,|M_2|\rightarrow\infty$, 
\beq
\delta{\cal L}=\frac{|e|H|m|}{2\pi}-\frac{1}{4\pi\sqrt{\pi}}
\int_{0+}^{\infty}\,\frac{d\eta}{\eta^{5/2}}\,e^{-m^2\eta}
\left[\frac{|e|H\eta\,(e^{-|e|H\eta}+|e|H\eta)}{{\rm sinh}(|e|H\eta)}
-1\right] \,,
\eeq
where a counterterm had been added in virtue of (30). The term 
proportional to $H^2$ in the series expansion corresponds to a charge 
renormalization finite counterterm 
\beqa
\delta{\cal L}_0=-\frac{e^2}{12\pi|m|}H^2 \,. 
\eeqa
      
\section{Scalar QED}

In s-QED, the energy levels of a charged boson in a constant and 
uniform magnetic field does not exhibit the spin degeneracy for 
$n\ne 0$. So, in four space-time dimensions, the negative vacuum 
zero-point energy turns out to be
\bea
-\varepsilon_0&=&\frac{|e|H}{(2\pi)^2}\int_{-\infty}^{\infty}\,dp_z\,
\sum_{n=0}^{\infty}\epsilon_{n}(p_z) \nonumber \\
&=&\frac{|e|H}{(2\pi)^2}\int_{-\infty}^{\infty}\,dp_z\sum_{n=0}^{\infty}
\sqrt{m^2+2|e|Hn+p_z^2}\,.
\eea
The regularized function to be evaluated is 
\beq
\Phi^R(H)=-\frac{\partial^2\varepsilon_0^R}{\partial (m^2)^2} 
=\sum_ic_i\Phi_i(H) \,,
\eeq
where
\bea
\Phi_i(H)&=&-\frac{|e|H}{2(2\pi)^2}\int_0^{\infty}\,dp_z\,
\sum_{n=0}^{\infty}(m_i^2+2|e|Hn+p_z^2)^{-3/2-\delta} \nonumber \\
&=&-\frac{|e|H}{2(2\pi)^2}\sum_{n=0}^{\infty}\frac{1}{m_i^2+2|e|Hn} \,.
\eea
As in section 2, the above momentum integrals were performed under 
conditions (11) for the regularized zero-point energy 
$\varepsilon_0^R$.

The functions $\Phi_i(H)$ can be integrated with the help of (16). In 
this case, the resulting correction for the electromagnetic Lagrangian 
density is given by
\bea
\delta{\cal L}&=&-\frac{1}{8\pi^2}\int_{0+}^{\infty}\,
\frac{d\eta}{\eta^3}\,e^{-m^2\eta}\left\{\frac{|e|H\eta}
{{\rm sinh}(|e|H\eta)}\left[e^{-|e|H\eta}+|e|H\eta+\dots\right]-1
\right\} \nonumber \\
&=&-\frac{1}{8\pi^2}\int_{0+}^{\infty}\,
\frac{d\eta}{\eta^3}\,e^{-m^2\eta}\,\left\{|e|H\eta{\rm coth}(|e|H\eta)
-1\right\} \,,
\eea
where both conditions (11) and (23) were applied to the regularized 
correction and {\em finite} counterterms had also to be added, so that 
parity invariance is retained. As expected, one ends up with (21), 
after charge renormalization. 

\section{Finite Temperature QED}

Finite temperature effects may also be included in the present discussion 
of effective field theories. For this purpose it is convenient to
consider, for example, the expectation value of the spinor QED
Hamiltonian in a thermal vacuum, as the one defined in Thermo Field 
Dynamics${}^{[18]}$. In this case, the correction to Maxwell's Lagrangian 
density reads
\beq
\delta{\cal L}=-[\varepsilon_0^{\beta}-(\varepsilon_0^{\beta})_
{{\bf H}=0}]\,\,,
\eeq
where
\beq
\varepsilon_0^{\beta}=\frac{|e|H}{(2\pi)^2}\sum_{n\sigma}\int_{-\infty}^{\infty}\,dp
\,\epsilon_{n\sigma}(p)
\left\{ \frac{1}{1+e^{\beta\epsilon_{n\sigma}}}-\frac{1}{1+e^{-\beta\epsilon_{n\sigma}}} 
\right\}\,.
\eeq
In the low temperature limit, $\beta\rightarrow\infty$, the integrand in 
the above momentum integral may be expanded as
\beq
\varepsilon_0^{\beta}=\varepsilon_0-\frac{2|e|H}{(2\pi)^2}\sum_{n\sigma}
\int_{-\infty}^{\infty}\,dp\,\epsilon_{n\sigma}(p)
e^{-2\beta\epsilon_{n\sigma}}\,.
\eeq 
The last expression can be evaluated in the case of a weak magnetic field $H$, 
by expanding the energy $\epsilon_n(p)$ for $n\neq 0$,
\[\varepsilon_n(p) \sim\sqrt{p^2+m^2}+\frac{|e|H}{\sqrt{p^2+m^2}}n\,\,,\]
so that (38) becomes
\beq
\varepsilon_0^{\beta} = \varepsilon_0 + \frac{2|e|H}{(2 \pi)^2} 
\frac{\partial}{\partial \beta} \int_0^{\infty}\,dp\,e^{- 2\beta \sqrt{p^2 + m^2}}
\left\{ 1 + 2\sum_{n=1}^{\infty} e^{-\frac{2 \beta |e|Hn}{\sqrt{p^2+m2}}}\right\}.
\eeq

Performing the sum in the last term of the above equation gives, in the limit 
$\beta \rightarrow \infty$,
\beq
\varepsilon_0^{\beta} = \varepsilon_0 + \frac{2|e|H}{(2 \pi)^2} 
\frac{\partial}{\partial \beta} \int_0^{\infty}\,dp\,\left\{ e^{- 2\beta \sqrt{p^2 + m^2}} + 
2e^{- 2\beta \sqrt{p^2 + m^2}} e^{-\frac{2 \beta |e|H}{\sqrt{p^2+m2}}}\right\}.
\eeq
Now, for a weak magnetic field, i.e., taking  
\[e^{-\frac{2 \beta |e|H}{\sqrt{p^2+m^2}}} \sim 1 - \frac{2\beta|e|H}{\sqrt{p^2 + m^2}}\,,\]
yields
\beq
\varepsilon_0^{\beta}=\varepsilon_0-\frac{3|e|H}{(2\pi)^2}\frac{\partial^2}{\partial \beta^2}
{\rm K}_0(2 \beta m)-\frac{8\beta|e|^2H^2}{(2\pi)^2}\frac{\partial}{\partial \beta}
{\rm K}_0(2 \beta m)\,,
\eeq
where
\beq
{\rm K}_0 = \int_{0}^{\infty}\,dp\,\, \frac{e^{-2 \beta \sqrt{p^2 + m^2}}}{\sqrt{p^2 + m^2}}\,
\eeq
is the zero order modified Bessel function. Finally, substituting (41) in (36),
\beq 
\delta {\cal L} = \delta {\cal L}_{ren} + \frac{12|e|Hm^2}{(2\pi)^2}\frac{\partial^2}
{\partial (2 \beta m)^2}
{\rm K}_0(2 \beta m)  + \frac{16 \beta m |e|^2H^2}{(2\pi)^2}\frac{\partial}
{\partial (2 \beta m)}
{\rm K}_0(2 \beta m).
\eeq
The same procedure applies to the scalar case, giving
\beq 
\delta {\cal L} = \delta {\cal L}_{ren} + \frac{16|e|Hm^2}{(2\pi)^2}\frac{\partial^2}
{\partial (2 \beta m)^2}
{\rm K}_0(2 \beta m) + \frac{16 \beta m |e|^2H^2}{(2\pi)^2}\frac{\partial}{\partial(2 \beta m)}
{\rm K}_0(2 \beta m).
\eeq

\section{Concluding Remarks}

In the previous sections, radiative corrections to the electromagnetic
Lagrangian densities in (3+1) and (2+1) dimensions were derived  
considering vacuum polarization effects induced by a constant and 
uniform electromagnetic background, which modify the zero-point 
energy of the quantized field of a relativistic charged particle. 
Divergent quantities were rendered finite from the very beginning and 
acquired an unambiguous mathematical meaning along each step of 
calculations. For this purpose, a consistent subtraction scheme had to
be developed, introducing auxiliary regulator masses in the sense of 
Pauli-Villars-Rayski regularization. 

As in QED, divergent physical quantities, such as the vacuum zero-point 
energy, must be regularized according to its singular degree. In four 
dimensions, the regulator masses satisfy relations (11), the same 
conditions that eliminate the quadratic and logarithmic divergences in 
the second-order vacuum polarization amplitude of QED, which has also 
to be regularized as a whole object, in order to preserve gauge 
invariance${}^{[4]}$. The finite fourth-order Feynman amplitude for 
Delbr\"uck scattering corresponds, in the low-energy limit, to the 
effective theory of Euler-Kockel-Heisenberg${}^{[13]}$, rederived in 
section 2.

In the three-dimensional case, the requirements fulfilled by the 
auxiliary masses, namely
\beqa
\sum_{i}c_i=0\,,\,\sum_{i}c_i|m_i|=0\,,
\eeqa 
agree with the corresponding constraints on the Pauli-Villars-Rayski 
regulators in the second-order vacuum polarization amplitude of 
QED$_3$, to account for the linear and logarithmic 
divergences${}^{[16]}$. In section 3, only the divergent sector of the
vacuum zero-point energy had to be regularized. The finite part lead 
to an odd-parity contribution to the effective Lagrangian density,
\beqa
{\cal L}_{\rm odd}=\frac{|em|H}{2\pi} \,,
\eeqa 
which corresponds to a topological Chern-Simons term, radiatively 
induced in QED$_3$${}^{[15]}$. If both the finite and divergent 
sectors had been treated in the same foot, condition (30) would have 
canceled the above mentioned contribution ${\cal L}_{\rm odd}$. As a 
result, the correction to the electromagnetic Lagrangian density would
be written as
\beq
\delta{\cal L}=-\frac{1}{4\pi\sqrt{\pi}}\int_{0+}^{\infty}\,
\frac{d\eta}{\eta^{5/2}}\,e^{-m^2\eta}\left[|e|H\eta
{\rm coth}(|e|H\eta)-1\right] \,.
\eeq 
This is the same expression derived by Redlich${}^{[17]}$, applying
Schwinger's formalism${}^{[3]}$, without a proper analysis of 
regularization constraints.   

Weisskopf's zero-point energy method was also 
applied to s-QED. It was shown that, in virtue of the absence of spin 
degeneracy in the energy of charged bosons, the additional condition 
(23), together with the requirement of parity invariance, had to be 
imposed on the regularized electromagnetic correction, eliminating 
odd-parity contributions to the resulting effective Lagrangian 
density, which coincides with expression (21), derived from the 
fermion vacuum energy.

In the last section, it was discussed the natural extension of 
Weisskopf´s formalism to finite temperature quantum electrodynamics, 
by considering the vacuum energy of Thermo Field Dynamics. At low 
temperatures, the Euler-Kockel-Heisenberg effective Lagrangian density 
factorizes from the corresponding momentum distributions, giving rise to 
ultraviolet finite temperature corrections. In both spinor and scalar cases, 
there appear odd-parity contribuctions which, togheter with higher order 
terms, exponentially vanish as the temperature approaches zero.

Finally, it´s noteworthy to point out that Weisskopf´s approach to 
effective Lagrangians might also shed light into recent discussions on
radiative corrections in quantum field theories, such as the controversy
on dinamical CPT violation and Lorentz symmetry breaking in the electroweak 
sector of a renormalizable extension of the Standard Model${}^{[19]}$. This 
will be the subject of a forthcoming paper.

\section*{Acknowledgements}

The author thanks the Funda\c{c}\~{a}o de Amparo \`{a} Pesquisa do Estado 
de S\~{a}o Paulo (FAPESP, Brazil) grant 2000/14758-2 and Spanish MCyT grant 
PB98-0693, for the partial financial support.   
        	
\section{References}

\begin{description}
\item[{[1]}] H. Euler and B. Kockel, Naturwiss. {\bf 23} 246 (1935); W. 
Heisenberg and H. Euler, Zeits. f\"ur Phys. {\bf 98} 714 (1936). 
\item[{[2]}] V. Weisskopf, Kgl. Danske Videnskab. Selskab {\bf 14} No.6 
(1936).
\item[{[3]}] J. Schwinger, Phys. Rev. {\bf 82} 664 (1951). 
\item[{[4]}] G. Rayski, Acta Phys. Polonica {\bf 9} 129 (1948);W. Pauli
and F. Villars, Rev. Mod. Phys. {\bf 21} 434 (1949).
\item[{[5]}] W. Dittrich and M. Reuter, Effective Lagragians in Quantum 
Electrodynamics, Lecture Notes in Physics 220, Springer-Verlag (1985).
\item[{[6]}] E. S. Fradkin, D. M. Gitman and Sh. M. Shvartsman, Quantum 
Electrodynamics with Unstable Vacuum, Springer-Verlag (1991). 
\item[{[7]}] M. Gell-Mann and M. L\'evy, Nuov. Cim. {\bf 16} 705 (1960); 
R. Delbourgo and M. D. Scadron, Mod. Phys. Lett. A {\bf 10} 251 (1995).
\item[{[8]}] Y. Nambu and G. Jona-Lasinio, Phys. Rev. {\bf 122} 345 
(1961); {\bf 124} 246 (1961); S. P. Klevansky, Rev. Mod. Phys. {\bf 64} 
3, 649 (1992).
\item[{[9]}] H. Georgi, Phys. Lett. B {\bf 240} 447 (1990).
\item[{[10]}] S. Deser, R. Jackiw and S. Templeton, Ann. Phys. (NY) {\bf
140} 372 (1982). 
\item[{[11]}] Kei-ichi Kondo, Nucl. Phys. B {\bf 450} 251 (1995); L. A. 
Manzoni, B. M. Pimentel and J. L. Tomazelli, Eur. Phys. J. C {\bf 8} 353
(1999); {\bf 12} 701 (2000). 
\item[{[12]}] W. Dittrich, Phys. Rev. D {\bf 19} 2385 (1979); P. Elmfors,
D. Persson e B. -S., Skagerstam, Phys. Rev. Lett. {\bf 71} 480 (1993);
I. J. R. Aitchison and C. D. Fosco, Phys. Rev. D {\bf 57} 1171 (1998).
\item[{[13]}] V. B. Berestetskii, E. M. Lifshitz and L. P. Pitaevskii, 
{\em Quantum Electrodynamics}, Butterworth and Heinemann, Second Edition
(1982).
\item[{[14]}] V. Fock, Physik. Z. Sowjetunion {\bf 12} 404 (1937).
\item[{[15]}] G. Scharf, W. F. Wreszinski, B. M. Pimentel and J. L. 
Tomazelli, Ann. Phys. {\bf 231} 1, 185 (1994).
\item[{[16]}] B. M. Pimentel and J. L. Tomazelli, Prog. Theor. Phys. 
{\bf 95} 1217 (1996).
\item[{[17]}] A. Redlich, Phys. Rev. D {\bf 29}, 2366 (1984).
\item[{[18]}] see J. L. Tomazelli and L. C. Costa, hep-th/0210031, and 
references therein.
\item[{[19]}] S. Carroll, G. Field and R. Jackiw, Phys. Rev. D {\bf 41},
1231 (1990); D. Colladay and V. A. Kosteleck\'y, Phys. Rev. D {\bf 58},
116002 (1998); J. M. Chung and B. K. Chung, Phys. Rev. D {\bf 63} 105015
(2001).
\end{description}

\end{document}